\documentclass[11pt,a4paper,aps,nofootinbib]{article}
\usepackage{jcappub}
\usepackage[utf8]{inputenc}
\usepackage{color}
\usepackage{amsmath, bm, physics}
\usepackage{amssymb, mathtools}
\usepackage{subcaption}
\usepackage{slashed}
\usepackage{appendix}

\usepackage{graphicx}
\usepackage{amsfonts}
\usepackage{leftindex}
\usepackage{color}
\usepackage{xcolor}
\usepackage{braket}
\usepackage{hyperref}
\usepackage{tikz}
\usetikzlibrary{arrows.meta}

\title{Cross-correlating Astrometric and Timing Residuals to Constrain Stochastic Gravitational-Wave Backgrounds}

\author[a,1]{E. Fink,\note{Corresponding author.}}
\author[a]{C.~R. Contaldi,}
\author[a]{and G. Mentasti}

\affiliation[a]{Blackett Laboratory, Imperial College London, SW7 2AZ, United Kingdom}

\emailAdd{ef513@cam.ac.uk}

\abstract{
We investigate the cross-correlation between astrometric and timing-residual observables for distant sources, such as pulsars and galaxies, and equivalent observables for nearby solar system bodies. Using the unified spin-weighted formalism introduced in \cite{spin1}, we derive the angular correlation functions—generalised Hellings–Downs curves—that describe the response of these mixed observables to a stochastic, unpolarised gravitational-wave background (SGWB). We compute the expected signal-to-noise ratio (SNR) and sensitivity for such measurements, focusing on cross-correlations between pulsar timing array (PTA) redshift signals and astrometric or distortion (shimmering) effects induced in solar system objects such as asteroids.  Although the current astrometric precision of asteroid tracking does not yet provide competitive constraints relative to PTA-only surveys, the method offers a complementary probe with enhanced sensitivity at higher frequencies. Future wide-field surveys capable of sub-milliarcsecond precision could make this approach a viable tool for detecting or constraining the SGWB. A key advantage of the technique is its reduced susceptibility to correlated systematics across different measurement domains, providing an independent cross-check of PTA detections and a potential observational bridge between PTA and LISA frequency bands.}

\begin{document}

\maketitle

\section{Introduction}
Since the first ground-based observation of a signal from a black-hole-binary merger at the Laser Interferometer Gravitational-Wave Observatory (LIGO) \cite{ligo}, gravitational waves (GWs) have become one of the most promising tools for discovering fundamental physics on astrophysical and cosmological scales. GWs are ripples, caused by the motion of massive objects, that perturb the spacetime metric and propagate at the speed of light.\footnote{This is the case in Einsteinian general relativity, but may not be in other gravity theories. \cite{subluminal}}

Current and future methods of observing GWs across the spectrum include ground-based interferometers such as LIGO at high frequencies $f \sim 100\;\text{Hz}$ \cite{ligo}, space-based interferometers such as the Laser Interferometer Space Antenna (LISA) at intermediate frequencies $f \sim 10^{-3}\;\text{Hz}$ \cite{lisa}, and Pulsar Timing Arrays (PTAs) at very low frequencies $f \sim 10^{-8}\;\text{Hz}$ \cite{pta}. 

Pulsars can be viewed as extremely precise clocks, as the variation in their rotational period is very small, leading to a corresponding small variation in the emission time of their pulses of EM radiation. PTAs utilise this by measuring the difference in arrival times between two consecutive pulses, shifted by a gravitational wave passing through the space between the pulsar and Earth. In addition to pulsars, the astrometric measurement of asteroids has also been explored as a viable source to measure GWs, as discussed in \cite{asteroids}. Although the relative accuracy of astrometric tracking is several orders of magnitude worse than that of timing measurements, the number of objects one could track is many orders of magnitude greater than the small number of pulsars being used in PTA observations. The complementarity in accuracy versus scale makes astrometric alternatives to PTA observations particularly interesting for constraining stochastic gravitational wave backgrounds (SGWB). Indeed, efforts along these lines are already underway using quasars \cite{darling} and galactic astrometry \cite{gaia-astrometry}.

A new generation of optical survey telescopes promises the accuracy needed to track solar system bodies, such as asteroids, making astrometric SGWB a possibility. For example, the Large Synoptic Survey Telescope (LSST) survey at the new Vera C. Rubin Observatory is expected to detect millions of small objects over the next 10 years \cite{vera-c-rubin}. The possibility of cross-correlating different low-frequency observables introduces two distinct advantages. The first is to measure the spectral properties of the time-dependent SGWB signal more robustly by combining observables that do not suffer correlated systematic uncertainties. The second is the ability to validate the nature of the SGWB using the expected angular correlations for each combination. In this work, we focus on the latter aspect and derive the expected angular correlation patterns induced by an SGWB in several combinations of observables.

This \emph{paper} is organised as follows. We make use of the formalism introduced in \cite{spin1} to write the cross-correlation of short-distance and long-distance observations in timing residuals, astrometry, and shimmering \cite{cosmic-shimmering}. In Section~\ref{obs}, we review the definition of spin-0 (scalar and pseudo-scalar), spin-1 (vector) and spin-2 (tensor) observables that capture the redshift, astrometry, and image distortion caused by an SGWB. In Sections~\ref{distortion-functions} and ~\ref{sec:corr}, we determine the angular two-point correlation functions between the combination of observables. The correlation functions are the analogues to the Hellings-Downs curve \cite{hellings-downs} for the generalised combinations. In Section~\ref{sec:SNR}, we present forecasts for the signal-to-noise ratio (SNR) and sensitivity curves that are expected when measuring the correlation functions. We end with a discussion of our findings and prospects for these new techniques in Section~\ref{sec:disc}.

\section{Astrometric, Shimmering, and Timing Observables}\label{obs}
To define our observables, we start by considering an image of a source with flux intensity $I(\hat{n})$ where $\hat{n}$ denotes the unit vector along the line-of-sight in angular direction $(\theta_s, \phi_s)$  on the sky. When GWs perturb the background null geodesic between an observer and the source, the image observed is distorted by
\begin{align}
    I_{obs}(\hat{n}) = I_{true}(\hat{n} + \delta \hat n)\;,
\end{align}
where $\delta \hat n$ is the distortion vector at any point on the sky \cite{cosmic-shimmering}. In the following, we introduce a notation $\delta\tilde{n}$ to specify the vector distortion induced by a GW of unit amplitude. This is motivated by our focus on the angular correlations induced by different observables. In this case, the angular response of the effect is of interest rather than the amplitude of the GW.\footnote{The discussion below follows the formalism introduced in detail in \cite{cosmic-shimmering,spin1} where the reader can follow a more in-depth introduction. Here, we focus on introducing the necessary concepts required for the particular application.}

When considering vector or tensor quantities, such as the astrometric and shimmering perturbations induced by distortion, one needs to define an orthonormal basis for decomposing such objects. We make use of an orthonormal basis on the plane perpendicular to the line-of-sight $\hat n$ or the direction of propagation of the GW $\hat{q}$. For example, when considering the direction $\hat{q} = (\theta, \phi)$, we define basis vectors perpendicular to $\hat q$ and lying on the tangent plane to it
\begin{subequations}
\begin{align}
    \hat{e}^\phi &= (-\sin{\phi}, \cos{\phi}, 0)\,, \\
    \hat{e}^\theta &= (\cos{\theta}\cos{\phi}, \cos{\theta}\sin{\phi}, -\sin{\theta})\,,
\end{align}
\end{subequations}
as well as the auxiliary vector
\begin{align}
    \hat{v}(\hat{q}) = \hat{e}^\theta + i\hat{e}^\phi\,,
\end{align}
that will be used to define a circularly polarised GW \cite{spin1}. For a GW aligned in the $\hat{q} = \hat{z}$ direction, this vector is
\begin{align}
    \hat{v}(\hat{z}) = e^{i\phi}(\hat{x} + i\hat{y})\,,
\end{align}
after a rotation of the arbitrary phase.
In standard general relativity, we have two Einsteinian transverse, traceless polarisations. The polarisation tensors defining these can be written in terms of the basis vectors introduced above as
\begin{align}
    \epsilon^+_{ij}(\hat{q}) = \hat{e}^\theta_i\hat{e}^\theta_j - \hat{e}^\phi_i\hat{e}^\phi_j\,, \\
    \epsilon^\times_{ij}(\hat{q}) = \hat{e}^\theta_i\hat{e}^\phi_j + \hat{e}^\phi_i\hat{e}^\theta_j\,.
\end{align}
We can also define left- and right-circularly polarised GWs as
\begin{align}
    \epsilon^{L/R}_{ij} = \epsilon^+_{ij} \pm i\epsilon^\times_{ij} \;.
\end{align}
In other theories of gravity with more degrees of freedom, there may also be scalar, vector and longitudinal polarisations or subluminal modes that will not be considered here \cite{asteroids,other-polarisations}.

In the basis $\{ \hat{e}^\theta, \hat{e}^\phi \}$, a left-circularly polarised GW travelling in a general $\hat{q}$ direction produces a metric perturbation of the form
\begin{align}
    h_{ij}(\hat{q}) \propto \epsilon^L_{ij}(\hat{q}) = [v\otimes v]_{ij}(\hat{q}) = v_iv_j \;,
\end{align}
which is the polarisation that we will use throughout this paper, but can be generalised easily \cite{cosmic-shimmering}.

Although we consider a perturbed metric here, indices should still be understood as raised and lowered with the Minkowski metric $\eta_{\mu\nu}$ in the $(-, +, +, +)$ convention, as we will only consider first-order effects. Specifically, this simplifies to the Euclidean metric $\delta_{ij}$ as we are only working with spatial dimensions here.

When considering the decomposition of observables at line-of-sight $\hat n$ it is convenient to align the tangential basis vectors with the geodesic on the sphere connecting the directions $\hat q$ and $\hat n$. This avoids unnecessary complications that would follow the introduction of a third, arbitrary reference direction such as $\hat z$. \footnote{This choice is particularly useful when considering statistically isotropic SGWBs where the orientation of the coordinate system should be irrelevant, but can lead to unnecessary algebraic complications.} This choice of basis about $\hat n$ can be defined as
\begin{subequations}
\begin{align}
    \hat{b}^\phi &= \frac{\hat{n}\times\hat{q}}{\sqrt{1-(\hat{n}\cdot\hat{q})^2}}\,, \\
    \hat{b}^\theta &= \frac{\hat{n}\times\hat{b}_\phi}{\sqrt{1 - (\hat{n}\cdot\hat{b}_\phi)^2}}\,,
\end{align}
\end{subequations}
and will be used to decompose astrometric (vector) and shimmering (tensor) observables  \cite{cosmic-shimmering} (see also \cite{mihaylov}). We can easily convert from any coordinate system to this basis using
\begin{subequations}
\begin{align}
    A_{ab} &= \hat{b}^i_a A_{ij} \hat{b}^j_b \\
    k_a &= \hat{b}^i_a k_i \;,
\end{align}
\end{subequations}
where $a, b$ run over the dimensions $\{ \hat{b}^\theta, \hat{b}^\phi \}$ of the plane perpendicular to the line of sight and $i, j$ run over the dimensions of an arbitrary coordinate system, in particular, Cartesian coordinates in this case.

Next, we define two spin-1 observables from the astrometric distortion $\delta\hat n$
\begin{align}
    \leftindex_{\pm}\delta = \delta \hat n \cdot (\hat{b}_\theta \pm i\hat{b}_\phi)\,,
\end{align}
with a response to a unit amplitude GW
\begin{align}
    F_{\pm\delta} = \delta\hat{n} \cdot (\hat{b}_\theta \pm i\hat{b}_\phi) \;,
\end{align}
where we note that $F_{+\delta}$ has spin $s = -1$ and $F_{-\delta}$ has spin $s = +1$ \cite{spin1}. Assuming the distortion effect is small, we can define the linear distortion matrix to describe the shimmering effect \cite{cosmic-shimmering}
\begin{align}
    \psi_{ab} = \frac{d(\delta \hat n)_a}{d\hat{n}^b} = -\kappa\delta_{ab} + \omega\varepsilon_{ab} + S_{ab}\,,
\end{align}
and the unit magnitude distortion matrix
\begin{align}
    \tilde{\psi}_{ab} = \frac{d(\delta \tilde{n})_a}{d\hat{n}^b} = -F_\kappa\delta_{ab} + F_\omega\varepsilon_{ab} + \tilde{S}_{ab} \;.
\end{align}
In the above, the $\kappa$ (scalar), $\omega$ (pseudo-scalar), and $S_{ab}$ (tensor) objects are the irreducible components of the general $\psi_{ab}$ tensor under rotations. In particular, $S_{ab}$ is the irreducible spin-2, symmetric, traceless component that can be represented using spin-weighted variables \cite{newman1966note}.

Using this decomposition  we define the spin-0 observables $\kappa$ and $\omega$ with response functions $F_\kappa$ and $F_\omega$ as well as the spin-2 observables
\begin{align}
    \leftindex_{\pm}\gamma = -(S_{11} \pm iS_{12})
\end{align}
with response functions
\begin{align}
    F_{\pm\gamma} = -(\tilde{S}_{11} \pm i\tilde{S}_{12}) \;,
\end{align}
where we note that $F_{+\gamma}$ has spin $s = 2$ and $F_{-\gamma}$ has spin $s = -2$ \cite{cosmic-shimmering}. In summary, we can write $\psi_{ab}$ in the $\{ \hat{b}_\theta, \hat{b}_\phi \}$ basis as
\begin{align}
    \psi = \begin{pmatrix}
                -\kappa - (\leftindex_+{\gamma} + \leftindex_-{\gamma})/2 & \omega + i(\leftindex_+{\gamma} - \leftindex_-{\gamma})/2 \\
                -\omega +i(\leftindex_+{\gamma} - \leftindex_-{\gamma})/2 & -\kappa +\leftindex_+{\gamma} + \leftindex_-{\gamma})/2
            \end{pmatrix}
\end{align}
which is the form we use to extract the observables and, equivalently, the response functions.

\section{Short-distance response functions}\label{distort}
Having defined the observables of interest, we now review the explicit expression of the distortion vector $\delta\hat{n}$ in the limit where the wavelength of the GW is much longer than the distance between the object and the observer, i.e. the short-distance limit appropriate for objects within the solar system, as discussed in \cite{asteroids}. As these objects appear point-like on a telescope, we will consider a group of them as an extended image, which allows us to define the distortion matrix. The general expression for the distortion vector, valid in any limit, was given by \cite{Book:2010pf,mihaylov} and the short-distance limit by \cite{asteroids} as
\begin{align}
    \delta {\hat n}^i = \frac{1}{2}h_{jk}{\hat n}^j(\delta^{ik} - {\hat n}^i{\hat n}^k) \;.
\end{align}
Following this, we can derive an expression for the distortion matrix
\begin{align}
    \psi_{ij} = \frac{1}{2}h_{jk}(\delta^k_i - {\hat n}_i{\hat n}^k) - \frac{1}{2}h_{rk}{\hat n}^r{\hat n}^k\delta_{ij} - \frac{1}{2}{\hat n}_ih_{jr}{\hat n}^r = h_{jk}(\frac{1}{2}\delta^k_i - {\hat n}_i{\hat n}^k) - \frac{1}{2}h_{rk}{\hat n}^r{\hat n}^k\delta_{ij} \;.
\end{align}
Considering a left-circularly polarised GW, defined in section \ref{obs}, the unit magnitude distortion vector and unit magnitude distortion matrix can be expressed as
\begin{align}
    \delta \tilde{n}^i &= \frac{1}{2}[v^i(\hat{n}\cdot\hat{v}) - {\hat n}^i(\hat{n}\cdot\hat{v})^2] \\
    \tilde{\psi}_{ij} &= \frac{1}{2}v_iv_j - {\hat n}_iv_j(\hat{n}\cdot\hat{v}) - \frac{1}{2}(\hat{n}\cdot\hat{v})^2\delta_{ij} \;.
\end{align}
We would like to compute the two-point correlation function between the short-distance observables introduced in Section~\ref{obs}, as measured from tracking of asteroids within the solar system, and the long-distance redshift effect, as measured from pulsar timing arrays, whose response function is given by (see, e.g. \cite{allen})
\begin{align}
    F_z = \frac{1}{2}\frac{(\hat{n}\cdot\hat{q})^2}{1+\hat{q}\cdot\hat{n}} \;.
\end{align} \\
Using the approach adopted in \cite{allen}, we consider a frame where the object being tracked or imaged is at $\hat{n} = (\theta_s, \phi_s)$ and the left-circularly polarised GW propagates in direction $\hat{z} = (\theta = 0, \phi)$. In this frame, the short-distance distortion observable response functions are
\begin{subequations}
\begin{align}
    F_\kappa(\hat{z}, \hat{n}) &= \frac{3}{4}e^{-2i(\phi-\phi_s)}\sin^2{\theta_s}\,, \\
    F_\omega(\hat{z}, \hat{n}) &= 0 \,, \\
    F_{+\delta}(\hat{z}, \hat{n}) &= \frac{1}{2}e^{-2i(\phi-\phi_s)}(\cos{\theta_s}+1)\sin{\theta_s} \,, \\
    F_{-\delta}(\hat{z}, \hat{n}) &= \frac{1}{2}e^{-2i(\phi-\phi_s)}(\cos{\theta_s}-1)\sin{\theta_s} \,, \\
    F_{+\gamma}(\hat{z}, \hat{n}) &= -e^{-2i(\phi-\phi_s)}\sin^4\left(\frac{\theta_s}{2}\right)\,, \\
    F_{-\gamma}(\hat{z}, \hat{n}) &= -e^{-2i(\phi-\phi_s)}\cos^4\left(\frac{\theta_s}{2}\right)\,, 
\end{align}
\end{subequations}
and the long-distance redshift response is given by
\begin{align}
    F_z(\hat{z}, \hat{n}) = \frac{1}{2}e^{2i(\phi-\phi_s)}(1-\cos{\theta_s}) \;.
\end{align}
As shown in \cite{cosmic-shimmering}, it is convenient to define scalar quantities using spin-raising and -lowering operators. This greatly simplifies the calculation of angular correlation functions because it removes the dependence of the coordinate frame under arbitrary rotations. To this end, we define spin raising ($\slashed{\partial}$) and lowering ($\bar{\slashed{\partial}}$) operators \cite{newman1966note,raising-lowering}
\begin{subequations}
\begin{align}
    \leftindex_{s+1}{p}(\theta, \phi) = \slashed{\partial} \leftindex_s{f}(\theta, \phi) = -\sin^s{(\theta)}\left( \frac{\partial}{\partial\theta} + i\csc{(\theta)}\frac{\partial}{\partial\phi}\right)\sin^{-s}{(\theta)}\leftindex_s{f}(\theta, \phi)\,, \\
    \leftindex_{s-1}{q}(\theta, \phi) = \bar{\slashed{\partial}} \leftindex_s{f}(\theta, \phi) = -\sin^{-s}{(\theta)}\left( \frac{\partial}{\partial\theta} - i\csc{(\theta)}\frac{\partial}{\partial\phi}\right)\sin^s{(\theta)}\leftindex_s{f}(\theta, \phi)\,.
\end{align}
\end{subequations}
These operators raise or lower the spin-weight of a function $\leftindex_s{f}(\theta, \phi)$ with spin $s$ and can be used to raise or lower the spin-weight of the response functions to spin-0 so that we can work with purely spin-0 objects that have simple, well-defined transformation properties \cite{raising-lowering}. These operators map $\bar{\slashed{\partial}}^2F_{+\gamma} = F_{+\zeta}, \; \slashed{\partial}^2F_{-\gamma} = F_{-\zeta}$ and $\bar{\slashed{\partial}}F_{-\delta} = F_{-\xi}, \; \slashed{\partial}F_{+\delta} = F_{+\xi}$ \cite{cosmic-shimmering} and give the spin-0 equivalent functions as 
\begin{subequations}
\begin{align}
    F_{\pm\zeta}(\hat{z}, \hat{n}) = -3e^{-2i(\phi-\phi_s)}\sin^2{\theta_s} \,,
     \\
    F_{\pm\xi}(\hat{z}, \hat{n}) = \frac{3}{2}e^{-2i(\phi-\phi_s)}\sin^2{\theta_s} \;.
\end{align}
\end{subequations}
We can now expand all response functions in terms of spin-0 (scalar) spherical harmonics $Y_{lm}(\theta_s, \phi_s)$. These restrict the expansion coefficients to $m = 2, \; l \geq 2$ due to the $e^{2i\phi_s}$ factor.

Determining the coefficients using overlap integrals and the orthonormality of spherical harmonics, we find that all short-distance functions give only quadrupole terms ($l = 2$)
\begin{align}
    F_X(\phi, \theta_s, \phi_s) = e^{-2i\phi}a^X_2 Y^\star_{22}(\theta_s, \phi_s) \;,
\end{align}
with coefficients are $a_2^\kappa = \sqrt{\frac{6\pi}{5}}$, $a_2^\omega = 0$, $a_2^\zeta = -4\sqrt{\frac{6\pi}{5}}$, and $a_2^\xi = 2\sqrt{\frac{6\pi}{5}}$. The coefficients for the long-distance redshift response $F_z$ are given by
\begin{equation}
    a^z_l = (-1)^l\sqrt{4\pi(2l+1)}\sqrt{\frac{(\ell+2)!}{(\ell-2)!}}\,,
\end{equation} 
and include contributions at all values of $l \geq 2$, although these will cancel later when taking the overlap integrals.

Having found the expansion of the distortion functions in terms of spherical harmonics, we use their well-defined transformation properties under rotation to find the distortion functions for a general GW direction defined by rotating the system such that $\hat z \to \hat{q}$
\begin{align}
    \bar Y_{lm}(\hat{n}) = \sum\limits_{m'} \mathcal{D}^l_{m'm}(\hat{q})Y_{lm'}(\hat{n}) \;,
\end{align}
where $\mathcal{D}^l_{m'm}$ are the Wigner D matrices that can be expressed as a function of Euler angles
\begin{align}
    \mathcal{D}^l_{m'm}(\hat q)\equiv\mathcal{D}^l_{m'm}(\phi, \theta, -\phi) = (-1)^{m'}\sqrt{\frac{4\pi}{2l+1}}\leftindex_m{Y}_{l, -m'}(\theta, \phi)e^{im\phi} \;,
\end{align}
where $\leftindex_m{Y}_{l, -m'}$ are spin-m weighted spherical harmonics \cite{spin-spherical-harmonics, gair-spin}. 

After rotating the functions, we have to invert the raising and lowering procedure to recover the original spin-1 and spin-2 observables. For this, we can use the identities
\begin{subequations}
\begin{align}
    (\bar{\slashed{\partial}})^{-1} Y_{2m} &= \frac{1}{\sqrt{6}}\leftindex_1{Y}_{2m},\; (\bar{\slashed{\partial}}^2)^{-1} Y_{2m} = \frac{1}{2\sqrt{6}}\leftindex_2{Y}_{2m}\,, \\
    (\slashed{\partial})^{-1} Y_{2m} &= -\frac{1}{\sqrt{6}}\leftindex_{-1}{Y}_{2m},\; (\slashed{\partial}^2)^{-1} Y_{2m} = \frac{1}{2\sqrt{6}}\leftindex_{-2}{Y}_{2m} \;.
\end{align}
\end{subequations}
Overall, the result of rotating and raising/lowering the distortion functions is summarised by the relations
\begin{subequations}
\begin{align}
    A^{s = 0}_2 &= \sqrt{\frac{4\pi}{5}}a_2\,,\\
    A^{s = \pm1}_2 &= \pm\frac{1}{\sqrt{6}}\sqrt{\frac{4\pi}{5}}a_2\,, \\
    A^{s = \pm2}_2 &= \frac{1}{2\sqrt{6}}\sqrt{\frac{4\pi}{5}}a_2 \;,
\end{align}
\end{subequations}
which can be used to express the general short-distance distortion response functions in terms of sums over mixed products of spin-weighted spherical harmonic basis functions
\begin{subequations}\label{distortion-functions}
\begin{align}
    F_\kappa(\hat{q}, \hat{n}) &= A_2^\kappa \sum\limits_{m = -2}^{2} \leftindex_2{Y}^{\,}_{2m}(\hat{q}) Y^\star_{2m}(\hat{n}) = \frac{2\sqrt{6}\pi}{5}\sum\limits_{m = -2}^{2} \leftindex_2{Y}^{\,}_{2m}(\hat{q}) Y^\star_{2m}(\hat{n})\,, \\
    F_\omega(\hat{q}, \hat{n}) &= 0 \,,\\
    F_{+\delta}(\hat{q}, \hat{n}) &= A_2^{+\delta}\sum\limits_{m = -2}^{2} \leftindex_2{Y}^{\,}_{2m}(\hat{q}) \leftindex_{-1}{Y}^\star_{2m}(\hat{n}) = -\frac{4\pi}{5}\sum\limits_{m = -2}^{2} \leftindex_2{Y}^{\,}_{2m}(\hat{q}) \leftindex_{-1}{Y}^\star_{2m}(\hat{n}) \,,\\
    F_{-\delta}(\hat{q}, \hat{n}) &= A_2^{-\delta}\sum\limits_{m = -2}^{2} \leftindex_2{Y}^{\,}_{2m}(\hat{q}) \leftindex_{1}{Y}^\star_{2m}(\hat{n}) = \frac{4\pi}{5}\sum\limits_{m = -2}^{2} \leftindex_2{Y}^{\,}_{2m}(\hat{q}) \leftindex_{1}{Y}^\star_{2m}(\hat{n}) \,,\,,\\
    F_{+\gamma}(\hat{q}, \hat{n}) &= A_2^{+\gamma}\sum\limits_{m = -2}^{2} \leftindex_2{Y}^{\,}_{2m}(\hat{q}) \leftindex_{2}{Y}^\star_{2m}(\hat{n}) = -\frac{4\pi}{5}\sum\limits_{m = -2}^{2} \leftindex_2{Y}^{\,}_{2m}(\hat{q}) \leftindex_{2}{Y}^\star_{2m}(\hat{n}) \,,\\
    F_{-\gamma}(\hat{q}, \hat{n}) &= A_2^{-\gamma}\sum\limits_{m = -2}^{2} \leftindex_2{Y}^{\,}_{2m}(\hat{q}) \leftindex_{-2}{Y}^\star_{2m}(\hat{n}) = -\frac{4\pi}{5}\sum\limits_{m = -2}^{2} \leftindex_2{Y}^{\,}_{2m}(\hat{q}) \leftindex_{-2}{Y}^\star_{2m}(\hat{n})\,.
\end{align}
\end{subequations}
The expressions above make explicit the spin dependence of the (spin-2) GW and that of observables with different spins (0, $\pm 1$, and $\pm 2$). Note also that the coordinate frame dependence of the observables with spin-$>0$ is now solely incorporated into the spin-weighted spherical harmonics, which are an explicit form of the generators of rotations on the sphere. 

The long-distance limit of the redshift response can also be written in this form \cite{allen}
\begin{align}
    F_z(\hat{q}, \hat{n}) &= \sum\limits_{l = 2} A_l^z \sum\limits_{m = -l}^l\leftindex_2{Y}_{lm}(\hat{q}) Y_{lm}(\hat{n}) = 4\pi\sum\limits_{l = 2}(-1)^l\sqrt{\frac{(\ell+2)!}{(\ell-2)!}} \sum\limits_{m = -l}^l \leftindex_2{Y}^{\,}_{lm}(\hat{q}) Y^\star_{lm}(\hat{n}) \;.
\end{align}

\section{Angular Correlation Functions}\label{sec:corr}
Having defined the angular response functions for all observables, we can now derive the angular correlation functions between long-distance redshift $z$ and short-distance observables. To do this, we consider an observable $A$ with spin $s_1$ at position $\hat{n}_1 = (\theta_1, \phi_1)$ and another object measuring observable $B$ with spin $s_2$ at position $\hat{n}_2 = (\theta_2, \phi_2)$.
An SGWB can be considered as a superposition of Fourier modes
\begin{align}
    h_{ij}(t, \mathbf{x}) = \sum\limits_P\int df \int d^2\hat{q}\;h_P(f, \hat{q}) \epsilon^P_{ij}(\hat{q})e^{2\pi if(t + \hat{q}\cdot\mathbf{x})}\,,
\end{align}
where $\epsilon^P_{ij}$ is the polarisation tensor for a polarisation $P$ defined in Section~\ref{obs}, and we are summing over left- and right-circular polarisations \cite{allen-otteweil}. The integrals are over the frequencies $f$ of the GWs in the stochastic spectrum as well as the GW source directions $\hat{q}$. In the limit where the many distant sources are adding incoherently to form the SGWB, we can assume stationarity and statistical isotropy, in which case the amplitude $h_P(f, \hat{q})$ for each polarization can be considered as a Gaussian random variable with
\begin{subequations}\label{stochastic-properties}
\begin{align}
    \braket{h_P(f, \hat{q})} &= 0\,, \\
    \braket{h_P(f, \hat{q})h^\star_{P'}(f', \hat{q}')} &= \delta_{PP'}\delta(f -f')\delta^2(\hat{q} - \hat{q}')H(f) \;,
\end{align}
\end{subequations}
where we have also assumed there is no net polarisation.

We now define two general observables
\begin{subequations}
\begin{align}
    A(\hat{n}_1) &= \sum\limits_P\int df \int d^2\hat{q}\,h_P(f, \hat{q}) F^P_A(\hat{q}, \hat{n}_1)(\hat{q})e^{2\pi if(t + \hat{q}\cdot\mathbf{x})}\,, \\
    B(\hat{n}_2) &= \sum\limits_P\int df \int d^2\hat{q}\,h_P(f, \hat{q}) F^P_B(\hat{q}, \hat{n}_2)(\hat{q})e^{2\pi if(t + \hat{q}\cdot\mathbf{x})} \;,
\end{align}
\end{subequations}
where $F^P_A(\hat{q}, \hat{n}_1)$ and $F^P_B(\hat{q}, \hat{n}_2)$ are the response functions for observables $A$ and $B$ respectively to a unit-magnitude GW with polarisation $P$.

To find the correlator between the observables, we insert the left-circularly polarised response functions $F^L_A(\hat{q}, \hat{n}_1)$ and $F^L_B(\hat{q}, \hat{n}_2)$, as stated in \eqref{distortion-functions}, and place the observer at the origin by setting $\mathbf{x} = 0$. As we will not be working with the right-circularly polarised response functions explicitly, we relabel $F^L_A(\hat{q}, \hat{n}_1)$ and $F^L_B(\hat{q}, \hat{n}_2)$ to $F_A(\hat{q}, \hat{n}_1)$ and $F_B(\hat{q}, \hat{n}_2)$. Ignoring the right-circularly polarised component, we first find the left-circularly polarised correlator, denoted by subscript $v$, 
\begin{align}
    \braket{A(\hat{n}_1, t)B^\star(\hat{n}_2, t')}_v &= \int df df' \int d^2\hat{q}d^2\hat{q}'\,\braket{h_L(f, \hat{q})h^\star_L(f', \hat{q}')} F_A(\hat{q}, \hat{n}_1)F^\star_B(\hat{q}', \hat{n}_2)(\hat{q})e^{2\pi i(ft-f't')}\,, \nonumber \\
    &= \int df H(f)e^{2\pi if(t-t')}\int d^2\hat{q}\, F_A(\hat{q}, \hat{n}_1) F^\star_B(\hat{q}, \hat{n}_2) \,, \nonumber \\
    &= \Gamma_v^{AB}(\hat{n}_1, \hat{n}_2)\int df H(f)e^{2\pi if(t-t')} \;,
\end{align}
where the frequency and angular components factorise when taking the correlator due to the isotropic property outlined in \eqref{stochastic-properties}, specifically as $H(f)$ does not depend on $\hat{q}$ \cite{spin1}.
We now focus on the angular correlation pattern $\Gamma_v^{AB}$ and introduce the \emph{unpolarised} correlation function as
\begin{align}\label{pattern}
    \Gamma^{AB}(\hat{n}_1, \hat{n}_2) = \frac{1}{2}[\Gamma^{AB}_v(\hat{n}_1, \hat{n}_2) + \Gamma^{\bar{B}\bar{A}}_v(\hat{n}_2, \hat{n}_1)]
\end{align}
where the notation $\bar{X}$ refers to the equivalent observable to $X$ with opposite spin for observables with non-zero spin and the same observable as $X$ for observables with zero spin \cite{spin1}.

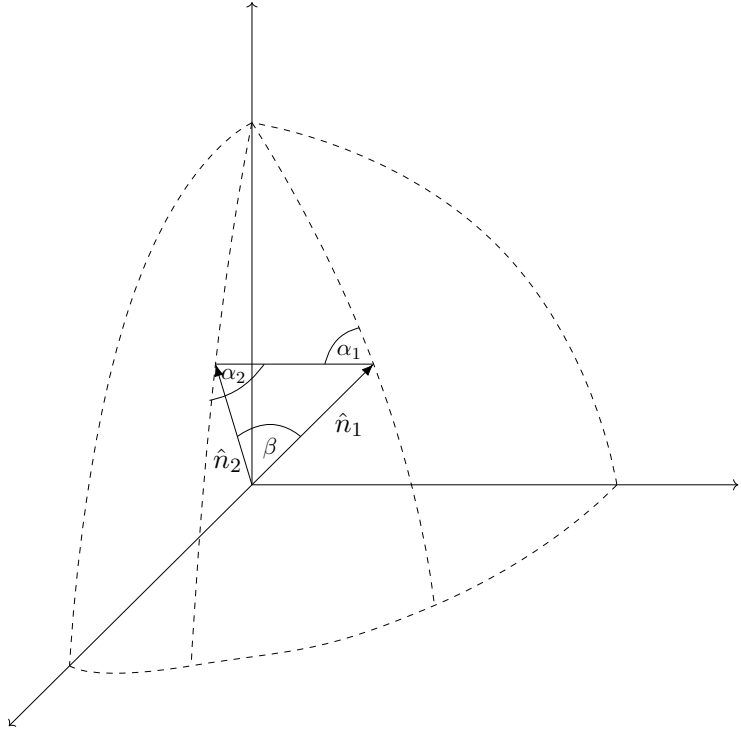
\begin{figure}[h]
    \centering
    \scalebox{0.8}{
    \begin{tikzpicture}
        \draw[->] (0, 0) -- (8, 0);
        \draw[->] (0, 0) -- (0, 8);
        \draw[->] (0, 0) -- (-4, -4);

        \draw[dashed] plot [smooth, tension = 1] coordinates {(0, 6) (-2, 3) (-3, -3)};
        \draw[dashed] plot [smooth, tension = 1] coordinates {(6, 0) (3, -2) (-1, -3) (-3, -3)};
        \draw[dashed] plot [smooth, tension = 1] coordinates {(6, 0) (4, 4) (0, 6)};
        \draw[dashed] plot [smooth, tension = 1] coordinates {(0, 6) (2, 2) (3, -2)};
        \draw[dashed] plot [smooth, tension = 1] coordinates {(0, 6) (-0.6, 2) (-1, -3)};

        \draw[-{Latex[length=2mm]}] (0, 0) -- (2, 2);
        \node[scale = 1.2] at (1.6, 1) {$\hat{n}_1$};
        \draw[-{Latex[length=2mm]}] (0, 0) -- (-0.6, 2);
        \node[scale = 1.2] at (-0.4, 0.4) {$\hat{n}_2$};
        \draw (2, 2) -- (-0.6, 2);

        \draw plot [smooth, tension = 1] coordinates {(0.8, 0.8) (0.3, 1) (-0.24, 0.8)};
        \node[scale = 1] at (0.3, 0.6) {$\beta$};
        \draw plot [smooth, tension = 1] coordinates {(1.2, 2) (1.4, 2.4) (1.76, 2.6)};
        \node[scale = 1] at (1.6, 2.2) {$\alpha_1$};
        \draw plot [smooth, tension = 1] coordinates {(0.2, 2) (-0.2, 1.6) (-0.7, 1.4)};
        \node[scale = 1] at (-0.3, 1.8) {$\alpha_2$};
    \end{tikzpicture}
    }
    \caption{Definition of angles on celestial sphere in  \eqref{spin-weight-addition}.}
    \label{fig:angles}
\end{figure}

For the left-circularly polarised correlation function, we have
\begin{align}
    \Gamma_v^{AB}(\hat{n}_1, \hat{n}_2) &= \int d^2\hat{q} F_A(\hat{q}, \hat{n}_1)F^\star_B(\hat{q}, \hat{n}_2)\,, \nonumber \\
    &= \int d^2 \hat{q} \left( \sum\limits_{lm} A^A_{l}\leftindex_2{Y}_{lm}(\hat{q})\leftindex_{s_1}{Y}^\star_{lm}(\hat{n}_1)\right) \left( \sum\limits_{l'm'} A^{B\star}_{l'} \leftindex_2{Y}^\star_{l'm'}(\hat{q})\leftindex_{s_2}{Y}_{l'm'}(\hat{n}_2) \right) \,, \nonumber \\
    &= \sum\limits_{lm}\sum\limits_{l'm'} A^A_{l}A^{B\star}_{l'}\leftindex_{s_1}{Y}^\star_{lm}(\hat{n}_1) \leftindex_{s_2}{Y}_{l'm'}(\hat{n}_2)\int d^2\hat{q} \leftindex_2{Y}_{lm}(\hat{q})\leftindex_2{Y}^\star_{l'm'}(\hat{q})\,,
\end{align}
which is the general form of the response functions as seen in Section~\ref{distort}. Using the completeness relations of spin-2 spherical harmonics \cite{varshalovich}, we obtain
\begin{align}
    \int d^2\hat{q} \leftindex_2{Y}_{lm}(\hat{q}) \leftindex_2{Y}_{l'm'}(\hat{q}) = \delta_{ll'}\delta_{mm'}\,,
\end{align}
such that the expression becomes
\begin{align}
    \Gamma_v^{AB}(\hat{n}_1, \hat{n}_2) &= \sum\limits_{lm} A^A_{l}A^{B\star}_{l} \leftindex_{s_1}{Y}^\star_{lm}(\hat{n}_1)\leftindex_{s_2}{Y}_{lm}(\hat{n}_2)\,, \nonumber \\
    &= \sum\limits_{l} A^A_{l}A^{B\star}_{l} \sum\limits_{m} \leftindex_{s_1}{Y}^\star_{lm}(\hat{n}_1)\leftindex_{s_2}{Y}_{lm}(\hat{n}_2) \;,
\end{align}
where we can apply the identity
\begin{align}\label{spin-weight-addition}
    \sum\limits_m \leftindex_{s_1}{Y}_{lm}(\hat{n}_1) \leftindex_{s_2}{Y}^\star_{lm}(\hat{n}_2) = \frac{2l+1}{4\pi}d^l_{s_1s_2}(\beta)e^{-i(s_1\alpha_1 + s_2\alpha_2)}\,,
\end{align}
where $\alpha_1, \;\alpha_2$ are the angles between the local meridians at the positions $\hat{n}_1, \;\hat{n}_2$ respectively and the geodesic connecting the directions on the surface of the celestial sphere and $\beta = \hat{n}_1 \cdot \hat{n}_2$ is the angle between the directions on the sky as shown in fig.~\ref{fig:angles} \cite{cosmic-shimmering, gair-spin}.

\begin{figure}[t]
    \centering
    \includegraphics[width=0.8\linewidth]{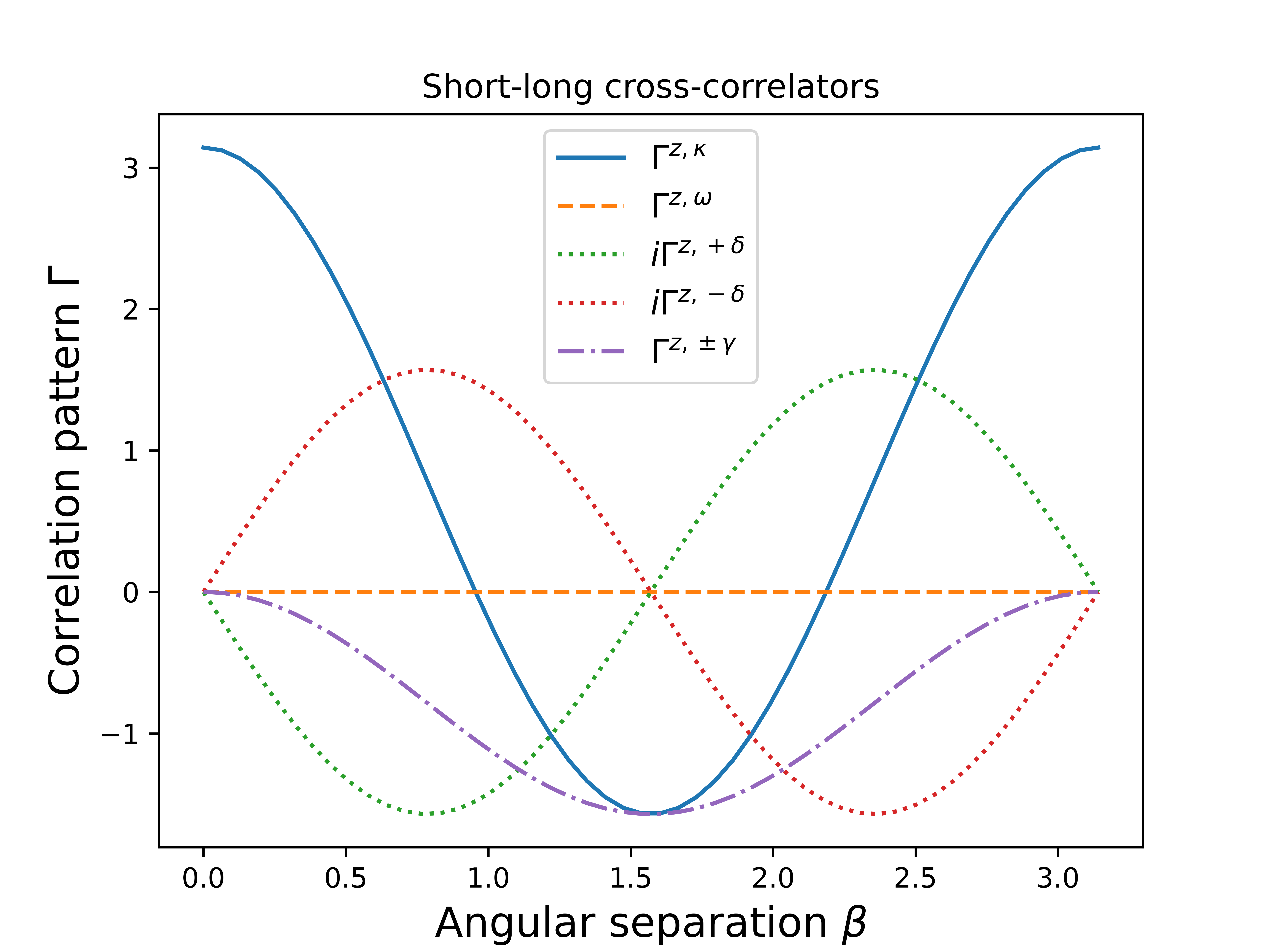}
    \caption{Plot of correlation patterns  \eqref{final-patterns} in the aligned frame where $\alpha = 0$ for $\Gamma^{z, \kappa}$, $\Gamma^{z, \omega}$, $\Gamma^{z, \pm\gamma}$ and $\alpha = \frac{\pi}{2}$ for $\Gamma^{z, \pm\delta}$.}
    \label{fig:correlators}
\end{figure}

We can now find the expression
\begin{align}
    \Gamma^{AB}_v(\beta) = \sum\limits_{l} A^A_{l}A^{B\star}_{l} \frac{2l+1}{4\pi}d^l_{s_1s_2}(\beta)e^{-i(s_1\alpha_1 + s_2\alpha_2)}\,,
\end{align}
where $d^l_{s_1s_2}(\beta)$ are the Wigner small-d matrices that form irreducible representations of $SO(3)$ \cite{smalld}. We can now substitute this into  \eqref{pattern}.
Applying the resulting expression to the coefficients from  \eqref{distortion-functions}, we can obtain the observable correlation patterns
\begin{subequations}\label{final-patterns}
\begin{align}
    \Gamma^{z, \kappa} &= \frac{5}{4\pi}\, A_2^\kappa A_2^z d^2_{00}(\beta) = \pi\,d^2_{00}(\beta) = \frac{\pi}{2}(3\cos^2{\beta} - 1)\,, \\
    \Gamma^{z, \omega} &= 0\,, \\
    \Gamma^{z, _\pm\delta} &= \frac{5}{4\pi}\,A_2^{_\pm\delta}A_2^zd^2_{0, \mp1}(\beta)(\pm i\sin{(\alpha_2)}) = -i\sqrt{\frac{2}{3}}\pi\,d^2_{0, \mp1}(\beta)\sin{(\alpha_2)} = \pm i\frac{\pi}{2}\sin(2\beta)\sin{(\alpha_2)}\,, \\
    \Gamma^{z, _\pm\gamma} &= \frac{5}{4\pi}\,A_2^{_\pm\gamma}A_2^zd^2_{0, \pm2}(\beta)\cos{(2\alpha_2)} = -\sqrt{\frac{2}{3}}\pi\,d^2_{0, \pm2}(\beta)\cos{(2\alpha_2)} = -\frac{\pi}{2}\sin^2{(\beta)}\cos{(2\alpha_2)}\,.
\end{align}
\end{subequations}
and plot these in a particular, aligned frame where we choose a value of $\alpha_2$. A natural choice is a frame where the spin $s \neq 0$ object is aligned along the $\hat{z}$ axis such that $\alpha_2 = 0$. However, this only works for the correlation patterns where $s_1 + s_2$ is even, as all unpolarised correlation patterns between observables where $s_1 + s_2$ is odd, such as $\Gamma^{z, \pm \delta}$, will vanish in the $\alpha_2 = 0$ frame. This is because they are related to a representation $\mathcal{D}^{s_1} \otimes \mathcal{D}^{s_2}$ of $SO(3)$ that transforms under parity as $\hat{P}(\mathcal{D}^{s_1} \otimes \mathcal{D}^{s_2}) = (-1)^{s_1 + s_2}(\mathcal{D}^{s_1} \otimes \mathcal{D}^{s_2}) = -(\mathcal{D}^{s_1} \otimes \mathcal{D}^{s_2})$, which leads to the contributions cancelling when the left and right circular contributions are summed. For this reason, we choose the $\alpha_2 = 0$ frame only for the correlators where $s_1 + s_2$ is even ($\Gamma^{z, \kappa}$, $\Gamma^{z, \omega}$ and $\Gamma^{z, \pm\gamma}$) and instead choose the $\alpha_2 = \frac{\pi}{2}$ frame for the remaining ones ($\Gamma^{z, +\delta}$ and $\Gamma^{z, -\delta}$). As they give imaginary values, we also scale these by a factor of $i$. This is shown in fig.~\ref{fig:correlators}.

Note that the dependence on the angle $\alpha_2$ and the associated ambiguity is not in conflict with our earlier assumption of statistical isotropy and is simply a consequence of the coordinate dependence of spin $>0$ variables. A more convenient way to state the correlations is to use the spin-weighted angular power spectra for each combination. These are coordinate invariant and any two observers would agree on their value irrespective of what choice of frame is made in the angular domain. In this case, they are pure quadrupoles
\begin{subequations}\label{Cls}
\begin{align}
C_2^{z, \kappa} &= A_2^\kappa A_2^z \,, \\
    C_2^{z, \omega} &= 0\,, \\
    C_2^{z, _\pm\delta} &= \frac{5}{4\pi}\,A_2^{_\pm\delta}A_2^z\,, \\
    C_2^{z, _\pm\gamma} &= \frac{5}{4\pi}\,A_2^{_\pm\gamma}A_2^z\,.
\end{align}
\end{subequations}

\section{Forecasts for Signal-to-Noise and Sensitivity}\label{sec:SNR}
 We now consider forecasts for the signal-to-noise ratio (SNR) of measuring each of the correlation patterns using LSST tracking of solar system asteroids and NANOGrav data for pulsars \cite{nanograv}.

The astrometric measurement of each object, labelled by $I$, will result in a time-domain data stream $d_I(t)$. This can be Fourier transformed into the frequency domain $\tilde{d}_I(f)$ and can be considered as a combination of signal and noise components\footnote{In practice, this transformation is complicated by inhomogeneous sampling of the object's positions, but we make the assumption that a simple Fourier transform will suffice for this initial exploration.}
\begin{align}
    \tilde{d}_I(f) = \tilde{h}_I(f) + \tilde{n}_I(f)\,,
\end{align}
with the GW signal $\tilde{h}_I(f)$ and noise $\tilde{n}_I(f)$ \cite{vaglio2025searching}. We assume stationarity in the noise, which means it can be characterised by a diagonal variance in the frequency domain
\begin{align}
    \braket{\tilde{n}_I(f)\tilde{n}_J(f')} = \Sigma_I(f)\delta(f - f')\delta_{IJ}\,,
\end{align}
for pairs of uncorrelated objects labelled by $I$ and $J$.

Following \cite{asteroids} and \cite{snr-allen}, we define the SNR as
\begin{align}
    \text{SNR}_{AB}^2 = \braket{|\Gamma^{AB}|^2}\sum\limits_{I, J} T \int\limits_{1/T}^{1/\Delta T} df \frac{H^2(f)}{\Sigma_I(f)\Sigma_J(f)} \;,
\end{align}
where we are summing over all pairs of objects that are being cross-correlated in the observables $A$ and $B$, $\Delta T$ is the cadence of observations, and $T$ is the total time of observation.

In our case, the $I$ index labels asteroids in the short-distance limit and the $J$ index labels pulsars in the long-distance limit. For the GWs, we assume a power-law spectrum
\begin{align}
    H(f) = \frac{3H_0^2}{32\pi^3f}\Omega_0\left( \frac{f}{f_{r}} \right)^{-\gamma} \;,
\end{align}
where $\gamma = 0$ for cosmological sources and $\gamma = \frac{13}{3}$ for astrophysical sources such as Supermassive Black Hole Binaries, $H_0$ is the Hubble parameter and $\Omega_0$ gives the amplitude of the GW background \cite{asteroids}. We also assume that the astrometric noise for the asteroids consists purely of a white spectrum
\begin{align}
\Sigma_{\rm ast} = \sigma_{\rm ast}^2\Delta T\,,
\end{align}
which means the noise is independent of frequency, whereas for pulsars, we have white and red noise components
\begin{align}
    \Sigma_{\rm pul} = 2f^2\left[\Sigma_{\rm pul}^{WN} + A^{RN}_{\rm pul}\left(\frac{f}{f^{RN}_r}\right)^{\gamma^{RN}}\right] \;,
\end{align}
where the white noise component is $\Sigma_{\rm pul}^{WN} = \tilde{\sigma}_{\rm pul}^2\Delta T$, similarly to the asteroid noise, and the new red noise component is assumed to follow a power law with $\gamma^{RN} = -3$ and all other parameters are determined empirically \cite{noise}. For the sake of these calculations, we use $A^{RN}_{\rm pul} = 6 \times 10^{-9}$ and $f^{RN}_r = \text{year}^{-1}$.

Starting with only long-long distance pulsars, we obtain the SNR
\begin{align}
    (\text{SNR}^{\rm pul}_{AB})^2 = \braket{|\Gamma^{AB}_{\rm pul}|^2}\frac{{\hat n}_{\rm pul}({\hat n}_{\rm pul}-1)}{2} T \int\limits_{1/T}^{1/\Delta T} df \frac{H^2(f)}{\Sigma_{\rm pul}(f)^2}\,,
\end{align}
which will help us to compare the cross-correlating method to a PTA only method later \cite{pta}.

Assuming all asteroids and pulsars have similar noise, we can arrive at the expression for the long-short distance cross-correlation SNR
\begin{align}\label{SNR}
    (\text{SNR}^{\rm cross}_{AB})^2 = \braket{|\Gamma^{AB}_{\rm cross}|^2}{\hat n}_{\rm pul}{\hat n}_{\rm ast} T \int\limits_{1/T}^{1/\Delta T} df \frac{H^2(f)}{\Sigma_{\rm pul}(f)\Sigma_{\rm ast}} \;.
\end{align}
Again, moving to the frame where $\alpha_2 = 0$ for correlators with $s_1 + s_2$ is even and $\alpha_2 = \frac{\pi}{2}$ for the remaining ones, we can compute the average correlation patterns as
\begin{subequations}
\begin{align}
    \braket{|\Gamma^{z, \kappa}(\beta)|^2} &= \frac{\pi^2}{5}\,, \\
    \braket{|\Gamma^{z, \omega}(\beta)|^2} &= 0\,, \\
    \braket{|\Gamma^{z, _\pm\delta}(\beta)|^2} &= \frac{2\pi^2}{15}\,,\\
    \braket{|\Gamma^{z, _\pm\gamma}(\beta)|^2} &= \frac{2\pi^2}{15}\;,
\end{align}
\end{subequations}
where all non-vanishing are of order 1, so we can treat them as equivalent to unity.

There is an important issue arising from the fact that, while grouped into an extended object, the asteroids themselves are point-like and not genuinely extended. For this reason, their light flux cannot be integrated properly, as would be the case for other astronomical objects that appear as genuinely extended, such as galaxies. The effect of this is that the SNR of the shimmering observables $\kappa$, $\omega$, and $\leftindex_{\pm}\gamma$ is not well-defined. While we still present the computations in principle, it is important to keep in mind that the following considerations are only strictly correct for the scalar observables $\leftindex_{\pm}\delta$.

We can now rewrite the SNR from  \eqref{SNR} in terms of the characteristic strain $h_c(f) = \sqrt{fH(f)}$ which follows a power law
\begin{align}\label{hc}
    h_c(f) = A_\alpha \left(\frac{f}{f_r}\right)^\alpha\,,
\end{align}
such that $\gamma = 2\alpha$ and $A_\alpha$ encodes the other constants \cite{contaldi-golat}. Hence, we can find the SNR in terms of the characteristic strain
\begin{align}\label{SNRhc}
    (\text{SNR}^{\rm cross})^2 = T {\hat n}_{\rm pul}{\hat n}_{\rm ast} \int\limits_{1/T}^{1/\Delta T} df \frac{h_c^4(f)}{f^2\Sigma_{\rm pul}(f)\Sigma_{\rm ast}}\,,
\end{align}
again assuming $\braket{|\Gamma^{AB}|^2} \approx 1$. This allows us to find the total SNR, setting $A_{\alpha} = 10^{-15}$ for a standard background. Our assumed survey parameters are $\tilde{\sigma}_{\rm pul} = 320 \;\text{ns}$, $\sigma_{\rm ast} = \sigma_{\rm LSST} = 50$ mas, $T = 5$ years, $\Delta T = 14$ days, ${\hat n}_{\rm pul} = 36$, ${\hat n}_{\rm ast} = 5 \times 10^{6}$ and $\alpha = 0$, which corresponds to a cosmological background \cite{contaldi-golat, asteroids}. This gives $\text{SNR}^{\rm pul} = 2.6$ and $\text{SNR}^{\rm cross} = 1.1 \times 10^{-4}$.

\begin{figure}[h]
    \centering
    \includegraphics[width=0.8\linewidth]{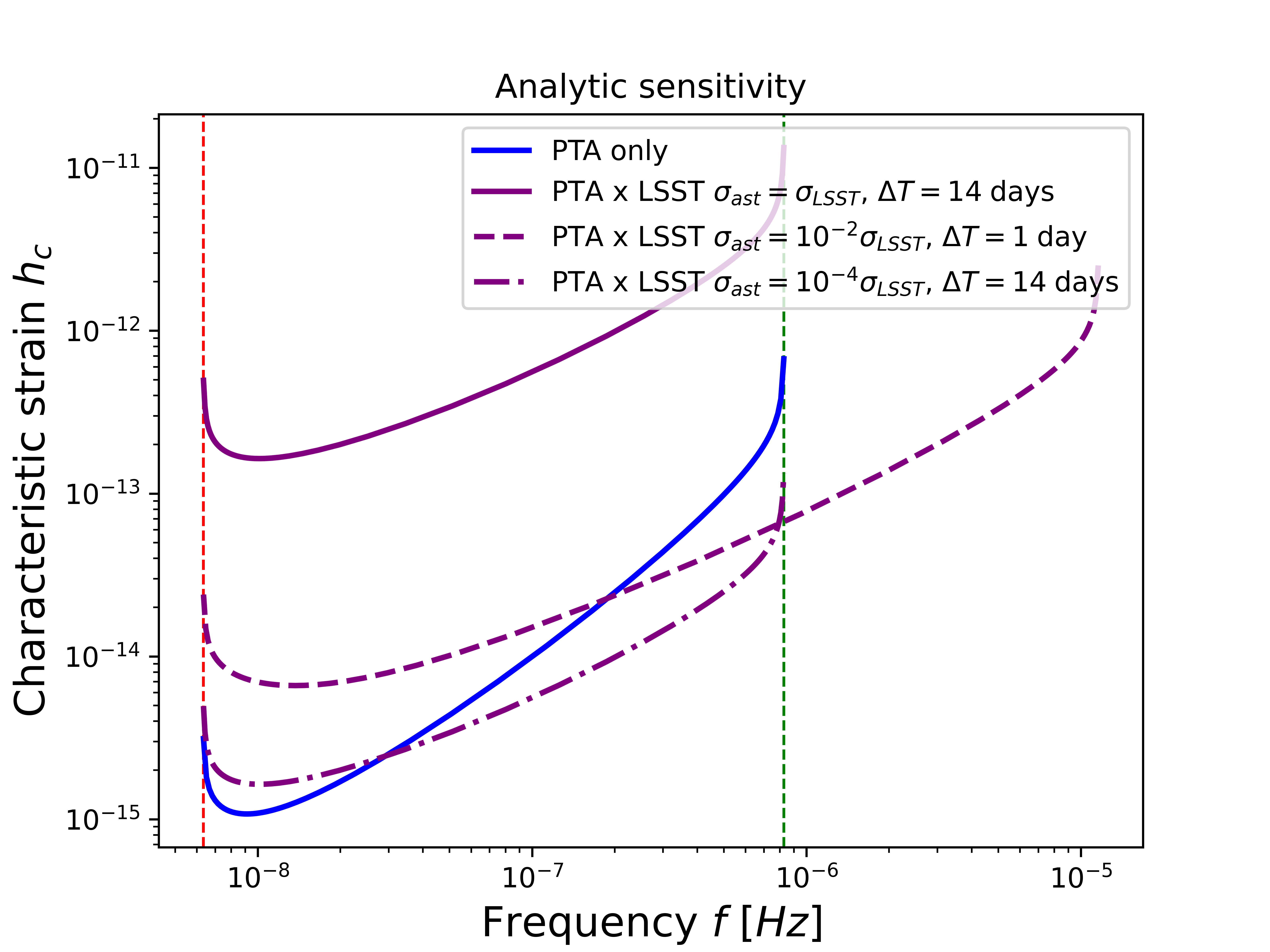}
    \caption{Analytic sensitivity for the cross-correlation between PTA and Vera C. Rubin (asteroids) according to  \eqref{SNR}, \eqref{SNRhc} compared to the sensitivity curve of only PTA measurements. Three cross-correlation curves are given for the state-of-the-art LSST survey, tracking asteroids with angular resolution $\sigma_{\rm ast} = \sigma_{\rm LSST}$ and cadence $\Delta T = 14$ days, and two potential future missions with $\sigma_{\rm ast} = 10^{-2}\sigma_{\rm LSST}$ and $\Delta T = 1$ day and $\sigma_{\rm ast} = 10^{-4}\sigma_{\rm LSST}$ and $\Delta T = 14$ days respectively. The vertical lines indicate the minimum frequency $\frac{1}{T} = \frac{1}{14\;\text{days}}$ (red) and maximum frequency $\frac{1}{\Delta T} = \frac{1}{1\;\text{day}}$ (green) that LSST is sensitive to.}
    \label{fig:contaldi-golat}
\end{figure}

Now we can find a sensitivity curve within the frequency range. We choose $\text{SNR} = 3$ as a detection limit as done in \cite{contaldi-golat} and find the coefficient $A_\alpha$ for different values of $\alpha$
\begin{align}\label{A-alpha}
    A_\alpha = \left(\frac{T {\hat n}_{\rm pul}{\hat n}_{\rm ast}}{9} \int\limits_{1/T}^{1/\Delta T} df \frac{\left(\frac{f}{f_r}\right)^{4\alpha}}{f^2\Sigma_{\rm pul}(f)\Sigma_{\rm ast}}\right)^{-\frac{1}{4}} \;,
\end{align}
where the integral is unfortunately not analytically solvable due to the red component of the pulsar noise. However, we can divide the frequency range into two sections where red and white noise dominate respectively, the threshold being denoted by $f_{\rm thresh}$. Neglecting the non-dominant contribution, the integral becomes analytically solvable, and the solution can be expressed as a sum. \\
We can now write $A_\alpha$ as
\begin{align}\label{A-alpha-analytic}
    A_\alpha \approx \left({\frac{T{\hat n}_{\rm ast}{\hat n}_{\rm pul}{}}{9\Sigma_{\rm ast}f^{4\alpha}_r}\left[\frac{\left[ \left(\frac{1}{\Delta T}\right)^{\lambda_{WN}} - (f_{\rm thresh})^{\lambda_{WN}}\right]}{\lambda_{WN}\Sigma^{WN}_{\rm pul}} + \frac{\left[\left(f_{\rm thresh}\right)^{\lambda_{RN}} - \left(\frac{1}{T}\right)^{\lambda_{RN} }\right]}{\lambda_{RN}(f^{RN}_r)^{-\gamma^{RN}}A^{RN}_{\rm pul}}\right]}\right)^{-\frac{1}{4}} \;,
\end{align}
where $\lambda_{WN} = 4\alpha - 5$ and $\lambda_{RN} = 4\alpha - \gamma^{RN} - 5$. For $\lambda_{WN} = 0$ or $\lambda_{RN} = 0$, $A_\alpha$ has a logarithmic dependence on either $\Delta T$ or $T$, but as we expect some continuous distribution of values for $\alpha$ in the stochastic background, we can safely ignore these case dues to their zero measures. This gives a family of parametric curves for $h_c(f, \alpha)$ by substituting  \eqref{A-alpha} or  \eqref{A-alpha-analytic} into  \eqref{hc} with an envelope as shown in fig. \ref{fig:contaldi-golat}. The envelope shows the minimum characteristic strain that the measurement technique is sensitive to at each frequency and hence provides a better way to compare different approaches than only using the total SNR.
 \\
We notice that the PTA only survey is better than the cross-correlation in the surveyed frequency range and hence provides a much better overall SNR, as seen above. The main reasons for this is that PTAs can be measured with a relatively small timing uncertainty of $100\;\text{ns}$ \cite{pta} whereas asteroids are tracked with an angular resolution of $50\;\text{mas}$ \cite{vera-c-rubin}, which is not ideal even compared to other current surveys such as GAIA that can operate with an angular resolution of $0.2$ mas \cite{gaia}. Additionally, asteroids are often in unstable orbits with many external forces acting on them that can distort the data strongly in certain frequency bands \cite{asteroids}.

Hence, we can consider how the variation of the survey parameters for asteroids would change the sensitivity and overall SNR of the cross-correlation method. In fig.~\ref{fig:contaldi-golat}, we can observe how the cross-correlation method make improvements on the PTA only method at high frequencies if asteroids could be tracked with a better angular resolution and cadence. This could be a promising approach to closing the gap between PTA surveys \cite{pta} and LISA \cite{lisa} in the spectrum of GWs, especially using measurements at lower cadence that can access more of these frequencies which is not necessarily possible in practice for PTAs due to technical limitations \cite{pta}.

\begin{figure}[th]
    \centering
    \includegraphics[width=0.8\linewidth]{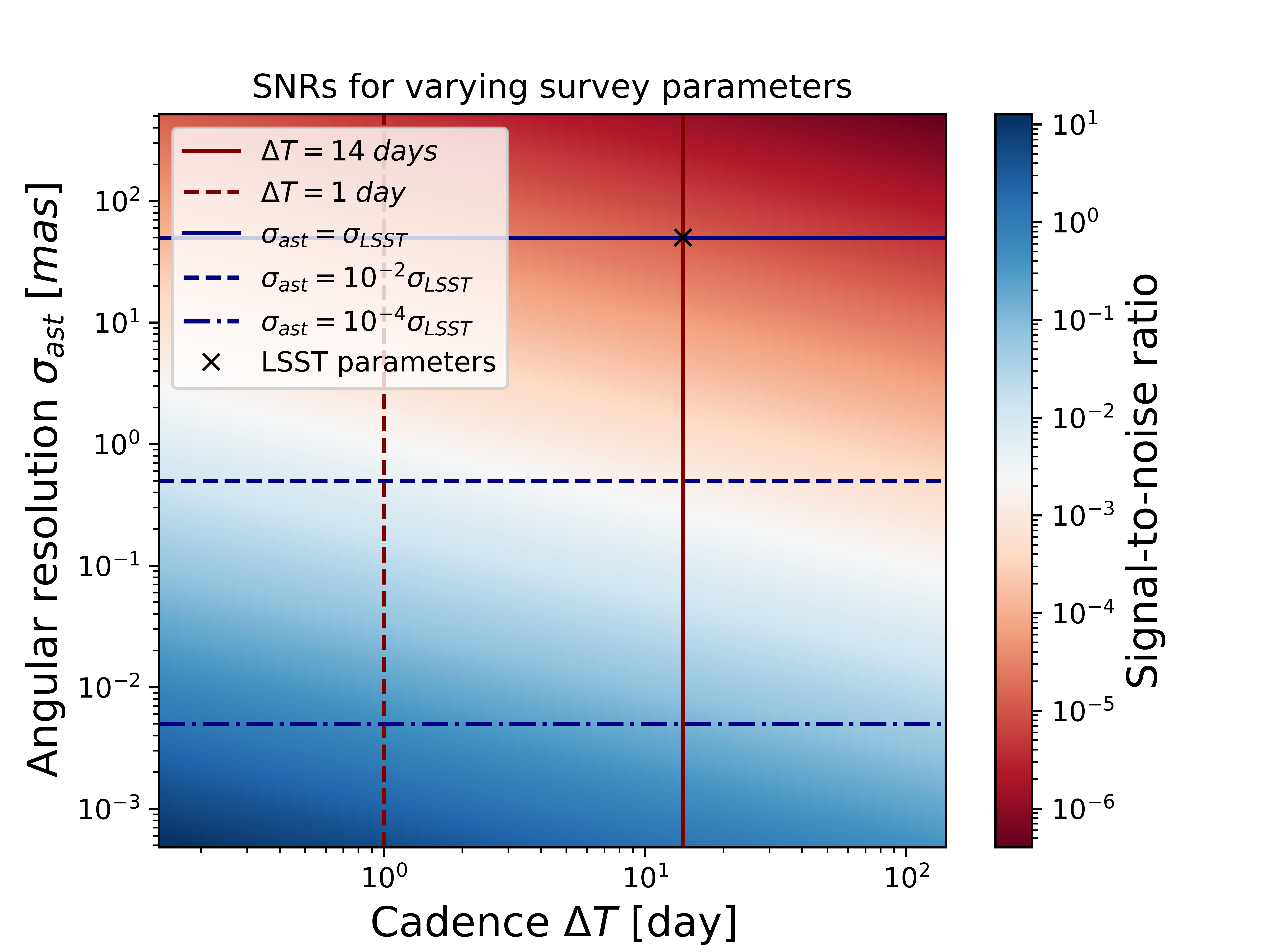}
    \caption{Variation of SNR with cadence and angular resolution of asteroid measurement. Different possible survey parameters are shown again, similarly to fig. \ref{fig:contaldi-golat}.}
    \label{fig:colourmap}
\end{figure}

Varying the angular resolution and cadence of the asteroids survey also has an effect on the total SNR, and hence the viability for detecting a stochastic GW background. This is shown in fig.~\ref{fig:colourmap}.

We can see that increasing the angular resolution of asteroid observations provides the most promising way for improving the overall SNR of the cross-correlation method. This could be technologically viable in the future since other stellar objects can already be tracked to such precision. Increasing the scanning frequency across the sky is another possible pathway to a better SNR although this could be less economical due to the need for a larger telescope or an array of telescopes.

\section{Discussion}\label{sec:disc}
In this work, we have defined scalar, vector, and tensor observables for solar system objects in the short-distance limit by grouping asteroids into effective extended images. Using these, we derived the unpolarised cross-correlation patterns between long-distance pulsar timing array (PTA) redshift observables and short-distance astrometric observables. The resulting long–short cross-correlation patterns simplify considerably: all scalar or scalar-equivalent observables are described by purely quadrupolar angular correlations, mirroring the structure of the standard Hellings–Downs–type correlations that appear in both long–long and short–short pairings.

With present astrometric precision, the achievable signal-to-noise ratio does not yet rival that of current PTA-only measurements. However, our analysis indicates that future surveys capable of higher angular resolution and cadence could make this cross-correlation technique a viable and complementary approach to probing stochastic gravitational-wave backgrounds (SGWBs). Improvements in asteroid tracking precision, in particular, could significantly enhance sensitivity at higher frequencies where PTA measurements lose efficacy.

A current limitation of the asteroid-grouping approach lies in estimating uncertainties for the shimmering observables, which cannot be straightforwardly adapted from analyses of genuinely extended sources such as galaxies. Since asteroid observations yield point-like data rather than integrated light profiles, our signal-to-noise forecasts apply most robustly to the scalar (spin-1) observables $\leftindex_{\pm}{\delta}$, while those involving spin-2 (shimmering) quantities require more careful error modelling in future work.

Looking ahead, two key advantages emerge from this long–short correlation method. First, by correlating high-cadence astrometric measurements with precise pulsar timing data, the technique could probe GW frequencies lying between the PTA and LISA sensitivity bands—potentially opening a new observational window on the SGWB spectrum. Second, because the two datasets originate from independent measurement systems—optical astrometry and radio timing—the cross-correlation is naturally robust against correlated instrumental systematics and common-mode noise. This property could provide a powerful means of verifying a stochastic GW signal once independent datasets of sufficient precision become available.

In summary, while current measurement capabilities limit the practical detectability of the predicted correlations, the formalism developed here establishes a clear theoretical framework and quantitative path for exploiting future astrometric–timing synergies in gravitational-wave cosmology.

\bibliographystyle{unsrt}
\bibliography{papers}

\begin{thebibliography}{10}

\bibitem{spin1}
Carlo~R Contaldi and Giorgio Mentasti.
\newblock A unified spin-harmonic framework for correlating pulsar timing, astrometric deflection, and shimmering gravitational wave observations.
\newblock {\em Preprint}, 2025.

\bibitem{ligo}
Benjamin~P Abbott, Richard Abbott, Thomas~D Abbott, Matthew~R Abernathy, Fausto Acernese, Kendall Ackley, Carl Adams, Thomas Adams, Paolo Addesso, Rana~X Adhikari, et~al.
\newblock Observation of gravitational waves from a binary black hole merger.
\newblock {\em Physical review letters}, 116(6):061102, 2016.

\bibitem{subluminal}
Wenzer Qin, Kimberly~K Boddy, and Marc Kamionkowski.
\newblock Subluminal stochastic gravitational waves in pulsar-timing arrays and astrometry.
\newblock {\em Physical Review D}, 103(2):024045, 2021.

\bibitem{lisa}
Scott~A Hughes.
\newblock Lisa sources and science.
\newblock {\em arXiv preprint arXiv:0711.0188}, 2007.

\bibitem{pta}
Michele Maiorano, Francesco De~Paolis, and Achille~A Nucita.
\newblock Principles of gravitational-wave detection with pulsar timing arrays.
\newblock {\em Symmetry}, 13(12):2418, 2021.

\bibitem{asteroids}
Giorgio Mentasti and Carlo~R Contaldi.
\newblock Observing gravitational waves with solar system astrometry.
\newblock {\em Journal of Cosmology and Astroparticle Physics}, 2024(05):028, 2024.

\bibitem{darling}
Jeremy Darling.
\newblock A new approach to the low-frequency stochastic gravitational-wave background: Constraints from quasars and the astrometric hellings--downs curve.
\newblock {\em The Astrophysical Journal Letters}, 982(2):L46, 2025.

\bibitem{gaia-astrometry}
Michael Perryman.
\newblock Space astrometry with gaia: Advances in understanding our galaxy.
\newblock {\em arXiv preprint arXiv:2509.10883}, 2025.

\bibitem{vera-c-rubin}
Colin~Orion Chandler, Pedro~H Bernardinelli, Mario Juri{\'c}, Devanshi Singh, Henry~H Hsieh, Ian Sullivan, R~Lynne Jones, Jacob~A Kurlander, Dmitrii Vavilov, Siegfried Eggl, et~al.
\newblock Nsf-doe vera c. rubin observatory observations of interstellar comet 3i/atlas (c/2025 n1).
\newblock {\em arXiv preprint arXiv:2507.13409}, 2025.

\bibitem{cosmic-shimmering}
Giorgio Mentasti and Carlo Contaldi.
\newblock Cosmic shimmering: the gravitational wave signal of time-resolved cosmic shear observations.
\newblock {\em Journal of Cosmology and Astroparticle Physics}, 2025(06):013, 2025.

\bibitem{hellings-downs}
RW~Hellings and GS~Downs.
\newblock Upper limits on the isotropic gravitational radiation background from pulsar timing analysis.
\newblock {\em Astrophysical Journal, Part 2-Letters to the Editor, vol. 265, Feb. 15, 1983, p. L39-L42.}, 265:L39--L42, 1983.

\bibitem{other-polarisations}
KJ~Lee, Fredrick~A Jenet, and Richard~H Price.
\newblock Pulsar timing as a probe of non-einsteinian polarizations of gravitational waves.
\newblock {\em The Astrophysical Journal}, 685(2):1304, 2008.

\bibitem{mihaylov}
Deyan~P Mihaylov, Christopher~J Moore, Jonathan~R Gair, Anthony Lasenby, and Gerard Gilmore.
\newblock Astrometric effects of gravitational wave backgrounds with non-einsteinian polarizations.
\newblock {\em Physical Review D}, 97(12):124058, 2018.

\bibitem{newman1966note}
Ezra~T Newman and Roger Penrose.
\newblock Note on the bondi-metzner-sachs group.
\newblock {\em Journal of Mathematical Physics}, 7(5):863--870, 1966.

\bibitem{Book:2010pf}
Laura~G. Book and Eanna~E. Flanagan.
\newblock {Astrometric Effects of a Stochastic Gravitational Wave Background}.
\newblock {\em Phys. Rev. D}, 83:024024, 2011.

\bibitem{allen}
Bruce Allen.
\newblock Pulsar timing array harmonic analysis and source angular correlations.
\newblock {\em Physical Review D}, 110(4):043043, 2024.

\bibitem{raising-lowering}
Matias Zaldarriaga and Uro{\v{s}} Seljak.
\newblock All-sky analysis of polarization in the microwave background.
\newblock {\em Physical Review D}, 55(4):1830, 1997.

\bibitem{spin-spherical-harmonics}
Kin-Wang Ng and Guo-Chin Liu.
\newblock Correlation functions of cmb anisotropy and polarization.
\newblock {\em International Journal of Modern Physics D}, 8(01):61--83, 1999.

\bibitem{gair-spin}
Jonathan~R Gair, Joseph~D Romano, and Stephen~R Taylor.
\newblock Mapping gravitational-wave backgrounds in modified theories of gravity using pulsar timing arrays.
\newblock {\em arXiv preprint arXiv:1506.08668}, 2015.

\bibitem{allen-otteweil}
Bruce Allen and Adrian~C Ottewill.
\newblock Detection of anisotropies in the gravitational-wave stochastic background.
\newblock {\em Physical Review D}, 56(2):545, 1997.

\bibitem{varshalovich}
Dmitri{\u\i}~Aleksandrovich Varshalovich, Anatol{\"\i}~Nikolaevitch Moskalev, and Valerij~Kel'manovǐc Khersonskii.
\newblock {\em Quantum theory of angular momentum}.
\newblock World Scientific, 1988.

\bibitem{smalld}
Jie Shen, Jie Xu, and Pingwen Zhang.
\newblock Approximations on so (3) by wigner d-matrix and applications.
\newblock {\em Journal of Scientific Computing}, 74(3):1706--1724, 2018.

\bibitem{nanograv}
Gabriella Agazie, Akash Anumarlapudi, Anne~M Archibald, Zaven Arzoumanian, Paul~T Baker, Bence B{\'e}csy, Laura Blecha, Adam Brazier, Paul~R Brook, Sarah Burke-Spolaor, et~al.
\newblock The nanograv 15 yr data set: evidence for a gravitational-wave background.
\newblock {\em The Astrophysical Journal Letters}, 951(1):L8, 2023.

\bibitem{vaglio2025searching}
Massimo Vaglio, Mikel Falxa, Giorgio Mentasti, Arianna~I Renzini, Adrien Kuntz, Enrico Barausse, Carlo Contaldi, and Alberto Sesana.
\newblock Searching for gravitational waves with gaia and its cross-correlation with pta: Absolute vs relative astrometry.
\newblock {\em arXiv preprint arXiv:2507.18593}, 2025.

\bibitem{snr-allen}
Bruce Allen and Adrian~C Ottewill.
\newblock Detection of anisotropies in the gravitational-wave stochastic background.
\newblock {\em Physical Review D}, 56(2):545, 1997.

\bibitem{noise}
Stanislav Babak, Mikel Falxa, Gabriele Franciolini, and Mauro Pieroni.
\newblock Forecasting the sensitivity of pulsar timing arrays to gravitational wave backgrounds.
\newblock {\em Physical Review D}, 110(6):063022, 2024.

\bibitem{contaldi-golat}
Sebastian Golat and Carlo~R Contaldi.
\newblock All-sky analysis of astrochronometric signals induced by gravitational waves.
\newblock {\em Physical Review D}, 105(6):063502, 2022.

\bibitem{gaia}
Antonella Vallenari, Anthony~GA Brown, Timo Prusti, Jos~HJ De~Bruijne, F~Arenou, Carine Babusiaux, Michael Biermann, Orlagh~L Creevey, Christine Ducourant, Dafydd~Wyn Evans, et~al.
\newblock Gaia data release 3-summary of the content and survey properties.
\newblock {\em Astronomy \& Astrophysics}, 674:A1, 2023.

\end{thebibliography}

\end{document}